\documentclass[10pt,twocolumn,letterpaper]{article}

\usepackage{cvpr}

\makeatletter
\@namedef{ver@everyshi.sty}{
\makeatother
\usepackage{tikz}}

\usepackage[T1]{fontenc}

\usepackage[utf8]{inputenc}

\usepackage{microtype}

\usepackage[accsupp]{axessibility}
\usepackage{amsmath}
\usepackage{amsfonts}
\usepackage{booktabs}
\usepackage{graphicx}
\usepackage[utf8]{inputenc}
\usepackage{makecell}
\usepackage{multirow}
\usepackage{xspace}

\usepackage{fontawesome5}

\usepackage{tabularx}
\usepackage{tikz}
\usetikzlibrary{calc}
\usetikzlibrary{backgrounds}

\newcommand{\aud}{\mathbf{a}}
\newcommand{\vis}{\mathbf{v}}
\newcommand{\txt}{\mathbf{t}}

\newcommand{\enc}{\mathrm{Enc}}
\newcommand{\encv}{\mathrm{Enc}_{\mathrm{v}}}
\newcommand{\dec}{\mathrm{Dec}}
\newcommand{\fuse}{\bowtie}

\newcommand{\method}[1]{\textsf{\small #1}}
\newcommand{\methodfs}[1]{\textsf{\footnotesize #1}}

\newcommand{\ia}{(i)\xspace}
\newcommand{\ib}{(ii)\xspace}
\newcommand{\ic}{(iii)\xspace}
\newcommand{\id}{(iv)\xspace}

\newcolumntype{Y}{>{\centering\arraybackslash}X}

\def\cvprPaperID{5} 
\def\confName{CVPR}
\def\confYear{2022}

\begin{document}

\title{Improving Multimodal Speech Recognition by Data Augmentation and Speech Representations}

\author{Dan Oneață \qquad Horia Cucu\\
    Speech and Dialogue Research Laboratory\\
    University \textsc{Politehnica} of Bucharest, Romania\\
    {\tt\small dan.oneata@speed.pub.ro}}

\maketitle

\begin{abstract}
    Multimodal speech recognition aims to improve the performance of automatic speech recognition (ASR) systems
    by leveraging additional visual information that is usually associated to the audio input.
    While previous approaches make crucial use of strong visual representations, \eg by finetuning pretrained image recognition networks,
    significantly less attention has been paid to its counterpart: the speech component.
    In this work, we investigate ways of improving the base speech recognition system
    by following similar techniques to the ones used for the visual encoder, namely, transferring representations and data augmentation.
    First, we show that starting from a pretrained ASR significantly improves the state-of-the-art performance;
    remarkably, even when building upon a strong unimodal system, we still find gains by including the visual modality.
    Second, we employ speech data augmentation techniques to encourage the multimodal system to attend to the visual stimuli.
    This technique replaces previously used word masking and comes with the benefits of being conceptually simpler and yielding consistent improvements in the multimodal setting.
    We provide empirical results on three multimodal datasets, including the newly introduced Localized Narratives.
\end{abstract}
%

\section{Introduction}
\label{sec:intro}

\begin{figure}
    \centering
    \begin{tikzpicture}[
            font=\footnotesize,
            block/.style={text width=1.25cm, align=center, minimum height=0.75cm},
            op/.style={circle, fill=blue!30, draw=blue!70, minimum size=0.7cm},
        ]
        \newcommand{\mylabel}[1]{{\color{black!50}\scriptsize\textsf{#1}}}
        \def\imgheight{0.75cm}

        \def\posa{0.5}
        \def\posb{2.30}
        \def\posc{4.00}
        \def\posd{5.70}
        \def\pose{7.70}

        \def\posx{0.9}
        \def\posy{-0.9}

        \newcommand\ab{\mathbf{a}}
        \newcommand\vb{\mathbf{v}}

        \node[label=\mylabel{audio}] at (\posa, \posx) (audio) {\includegraphics[height=\imgheight]{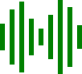}};
        \node[label=\mylabel{image}] at (\posa, \posy) (image) {\includegraphics[height=\imgheight]{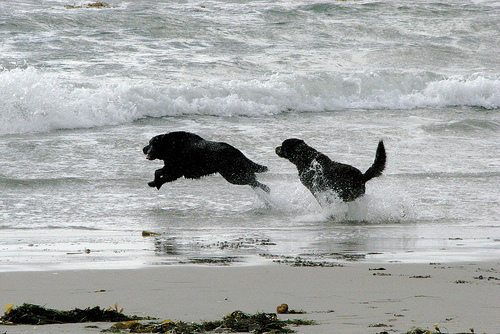}};

        \node[block, fill=red!15]    at (\posb,  \posx) (enc-audio) {audio encoder};
        \node[block, fill=orange!20] at (\posb,  \posy) (enc-image) {visual encoder};

        \node[op] at (\posc,  \posx) (concat) {$\bowtie$};

        \node[block, fill=red!15]           at (\posd,  \posx) (decoder) {decoder};
        \node[block, label=\mylabel{text}]  at (\pose,  \posx) (text)    {\it two dogs are playing in the sea};

        \draw[->] (audio) -- (enc-audio);
        \draw[->] (image) -- (enc-image);
        \draw[->] (enc-audio) -- (concat);
        \draw[->] (enc-image) -| (concat);
        \draw[->] (concat) -- (decoder);
        \draw[->] (decoder) -- (text);
    \end{tikzpicture}
    \caption{%
        Overview of a multimodal speech recognition system.
        Compared to traditional speech recognition, which generates transcriptions based solely on audio input,
        the multimodal setting involves using an additional input corresponding to the visual stream (\eg, image or video).
        The main motivation of this setup is that the visual information is often associated with audio (as encountered in instructional videos, documentaries, movies)
        and it can help disambiguate the audio recording, consequently, producing more accurate transcriptions.
        In this paper, we investigate ways of improving the multimodal system by focusing on the audio encoder and its fusion with the visual encoder.
    }
    \label{fig:overview}
\end{figure}
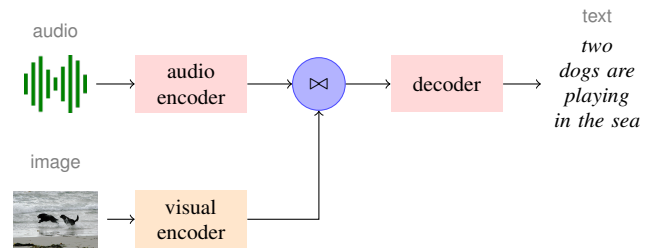

With the advent of video sharing platforms (such as YouTube or Vimeo),
multimodal data involving audio, visual and language are becoming ubiquitous.
In many types of video, such as instructional videos, documentaries, movies, what is spoken is related (grounded) to the visual channel.
In this paper we build upon this observation and address the task of automatic speech recognition in the context of visual information, also known as multimodal speech recognition.
Concretely, we assume that we have two inputs (the acoustic signal and a related visual modality, such as a video or an image) and we want to output the transcription of the input utterance.
The setup is illustrated in Figure \ref{fig:overview}.

The recent work on multimodal speech recognition makes crucial use of deep end-to-end architectures %
\cite{palaskar2018icassp,caglayan2019icassp,srinivasan2020icassp,srinivasan2020aemnlp,srinivasan2020bemnlp,paraskevopoulos2020acl,ghorbani2021slt}.
We follow their suite and develop an end-to-end multimodal speech recognition system.
Compared to previous work, we first experiment with two fusion mechanisms for combining the audio and visual modalities---%
either concatenation along the embedding dimension or concatenation along the temporal dimension.
Second, and more importantly, we improve the pipeline by focusing on the speech component by
\ia transferring pretrained speech representations and
\ib performing audio-level data augmentations.

\textbf{Transferring representations.}
All recent papers on multimodal speech recognition transfer visual representations, obtained as activations or softmax predictions of a pretrained visual classification network.
Depending on the training classes (%
objects, as in \cite{sun2016slt,moriya2018slt,palaskar2018icassp,srinivasan2020aemnlp};
scenes, as in \cite{miao2016interspeech,gupta2017icassp,srinivasan2020icassp};
actions, as in \cite{miao2016interspeech,caglayan2019icassp,paraskevopoulos2020acl};
faces, as in \cite{miao2016interspeech,moriya2019icassp}),
the visual encoder is more sensible to pick up certain types of visual information.
However, none of these prior works make use of pretrained speech representations.
In this paper we not only show the importance of starting from a good representation for both the audio and visual channels,
but crucially we provide an answer to question of whether the visual information is helpful for a stronger baseline system.

\textbf{Data augmentation.}
Augmenting the training set with perturbed samples is a common technique to enforce invariants for high-capacity deep learning models.
For image classification, images are altered by horizontal flips and small affine transformations, while for speech recognition the speed of an utterance is modified by time warping.
We use these ideas, in particular those related to speech augmentation, to improve the multimodal models.
Our intuition is that perturbing the audio signal will make the model more reliant on the visual channel.
The inspiration stems from the work of Srinivasan \etal \cite{srinivasan2019icml,srinivasan2020aemnlp} which have shown
that multimodal models improve over the baseline ASR even when audio-image pairs are mismatched (incongruent) \cite{srinivasan2019icml},
but if the multimodal models were trained on masked audio signals, this behaviour is alleviated \cite{srinivasan2020icassp}.
Compared to the previous approaches \cite{srinivasan2019icml,srinivasan2020icassp,srinivasan2020aemnlp}, we do not limit ourselves to temporal masking of words,
but randomly mask temporal and frequency segments, as in \cite{park2019interspeech}; as a consequence our approach is more general and convenient to use.
Another key distinction is that we do not carry the evaluation on the masked data,
but consider the more realistic scenario of assuming clean speech at test time and performing alterations only at train time.

Apart from the two main contributions of transfer learning and data augmentation,
we also explore fusion techniques for the audio and visual modalities.
Concretely, we propose two options: fusing the embeddings a) along the feature dimension and b) along the temporal dimension.
All these settings are empirically evaluated on three multimodal speech recognition datasets.
We report state-of-the-art performance on the commonly-used Flickr8K and How2 datasets, and new results on the recently introduced Localized Narratives dataset.


\section{Related work}
\label{sec:related-work}

In this section we discuss the main categories of multimodal models and present our task in the context of related problems.

\textbf{A taxonomy of multimodal models.}
Perhaps unsurprisingly, the techniques for multimodal speech recognition have been following the trends in speech processing and computer vision.
Based on the choices of the two main components (namely, the audio and visual pipelines), we distinguish three types of systems.

The first approaches \cite{mukherjee2003eurospeech,fleischman2008acl} date back to the 2000s and
rely on the Hidden Markov Models and Gaussian Mixture Models (the HMM-GMM paradigm) for speech recognition and hand-crafted features for the visual channel.
These methods also assumed more constrained and simplified settings to account for the lack of data.

The second category of multimodal systems \cite{miao2016interspeech,gupta2017icassp,sun2016slt,moriya2018slt,oneata2021cee} uses Hidden Markov Models and Deep Neural Networks (the HMM-DNN paradigm) for speech recognition,
while the visual component relies on pretrained networks.
While many of these approaches fuse the two components at the last stage (language modeling) \cite{gupta2017icassp,sun2016slt,moriya2018slt,oneata2021cee},
a notable exception is the work of Miao \etal \cite{miao2016interspeech}, which advocates for early fusion, at the audio level,
based on the observation that the acoustic conditions can correlate with the visual context.

Finally, the latest type of models leverage recent developments in end-to-end architectures and training \cite{palaskar2018icassp,caglayan2019icassp,srinivasan2020icassp,srinivasan2020aemnlp,srinivasan2020bemnlp,paraskevopoulos2020acl,ghorbani2021slt}.
For the audio part, the most common model involves recurrent networks for encoder and decoders, coupled through an attention mechanism,
but other variants include using a connectionist temporal classification (CTC) model (as done in \cite{palaskar2018icassp}) or the Transformer architecture, which involves attention-only layers (as done in \cite{paraskevopoulos2020acl}).
Various fusion levels have been explored:
encoder, decoder, and also at acoustic level.
Of course these can be combined as done by Caglayan \etal \cite{caglayan2019icassp}.

Our approach falls into the latter category, of end-to-end architectures.
We share similarities to the work of Paraskevopoulos \etal \cite{paraskevopoulos2020acl},
in that we employ Transformer architecture and sub-word modeling,
however our base speech recognition system is much stronger and we focus our empirical evaluation on the importance of transfered representations.

\textbf{Related tasks.}
We distinguish our work from two closely related tasks, which also make use of audio and visual input modalities.
A first task is audio-visual speech recognition \cite{mroueh2015icassp,petridis2018arxiv,afouras2018pami,ma2021icassp}, which also attempts to improve speech recognition, but it uses lip movement information.
A key difference to our methodology is that for the lip-based recognition there is a much tighter (although arguably more difficult to model) relation between the video and the transcriptions,
while for multimodal speech recognition, the relationship is at a semantic level and might affect only a few words, which have visual grounding.
Among the methods applied for this problem, the closest to our approach is the parallel work of Ma \etal \cite{ma2021icassp}, which also employs pretrained audio representations and data augmentations.
A second related task is learning audio-visual correspondences, but without depending on the textual annotations.
This formulation has the advantage of relying on less supervision and finds many uses, such as
representation learning \cite{harwath2016nips,harwath2015asru},
learning linguistic units \cite{harwath2019iclr},
semantic keyword spotting \cite{kamper2019taslp},
speech-based image retrieval \cite{synnaeve2014nips,harwath2016nips,harwath2018eccv} and
speech-based object localization \cite{harwath2018eccv}.






\section{Methodology}

\begin{figure*}
    \resizebox{\textwidth}{!}{
    \centering
    \begin{tikzpicture}[%
            font=\footnotesize,
            op/.style={circle, fill=blue!30, draw=blue!70, minimum size=0.7cm},
            data/.style={color=black!75, font=\sf\scriptsize, align=center},
            fork/.style={circle, fill, inner sep=1.5pt},
            dims/.style={font=\tiny, gray, scale=0.8},
        ]
        \def\posa{0.5}
        \def\posb{2}
        \def\posbc{2.5}
        \def\posc{3}
        \def\posd{4}
        \def\posde{5}
        \def\pose{5.5}
        \def\posf{7}
        \def\posg{8}
        \def\posh{9.5}

        \def\posx{0.9}
        \def\posy{-0.9}

        \begin{scope}
            \node[minimum height=1cm, minimum width=1.5cm, fill=red!15, label={\faVolumeUp}] at (\posa,  \posx) (speech-input-1) {};
            \node[data] at (\posa, \posx) (speech-input) {speech\\embeddings};
            \node[dims] at (\posa, 0.5) (speech-input-dims) {$D_{\aud} \times T$};

            \node[minimum height=1cm, minimum width=0.2cm, fill=orange!20, label={\faImage}] at (\posa,  \posy) (visual-input-1) {};
            \node[data] at (\posa, \posy) (visual-input) {image\\embedding};
            \node[dims] at (\posa, -1.3) (visual-input-dims) {$D_{\vis} \times 1$};

            \node[fork] at (\posb,  \posx) (speech-clone) {};

            \node[op] at (\posb, \posy) (visual-pi) {$\pi_D$};
            \node[op] at (\posc, \posy) (visual-rho)  {$\rho_T$};

            \node[op] at (\posc,  \posx) (speech-pi) {$\pi_D$};

            \node[op] at (\posd,  \posx) (concat) {$\bowtie_D$};

            \node[minimum height=1.5cm, minimum width=1.5cm, fill=red!15] at (\pose,  \posx) (speech-visual-1) {};
            \node[minimum height=0.75cm, minimum width=1.5cm, fill=orange!20] at (\pose,  0.54) (speech-visual-2) {};
            \node[data] at (\pose,  \posx) (speech-visual) {concat.\\embeddings};
            \node[dims] at (\pose, 0.5) (concat-dims) {$(D_{\aud'} + D_{\vis'}) \times T$};

            \node[op] at (\posf,  \posx) (speech-visual-pi) {$\pi_D$};
            \node[op] at (\posg,  \posx) (add) {$+$};

            \node[minimum height=1cm, minimum width=1.5cm, fill=red!15] at (\posh,  \posx) (output-1) {};
            \node[data] at (\posh,  \posx) (output) {output\\embeddings};
            \node[dims] at (\posh, 0.5) (output-dims) {$D_{\aud} \times T$};

            \draw [->] (speech-input) -- (speech-pi);
            \draw [->] (speech-pi)  -- (concat);
            \draw [->] (visual-input) -- (visual-pi);
            \draw [->] (visual-pi) -- (visual-rho);
            \draw [->] (visual-rho) -| (concat);
            \draw [->] (speech-clone) |- ($(concat.north) + (0.0, 0.7)$)  -| (add);
            \draw      (concat) -- (speech-visual);
            \draw [->] (speech-visual) -- (speech-visual-pi);
            \draw [->] (speech-visual-pi) -- (add);
            \draw      (add) -- (output);

            \begin{pgfonlayer}{background}
                \path (visual-input-1.south -| speech-input-1.west) + (-0.1, -0.1) node (bottom left) {};
                \path (concat.north -| output-1.east) + (0.1, 1.1) node (top right) {};
                \filldraw [gray!5] (top right) rectangle (bottom left);
                \path (visual-input-1.south -| output-1.east) + (-0.4, 0.2) node[blue!50!gray] (bottom right) {\method{emb}};
            \end{pgfonlayer}
        \end{scope}

        \begin{scope}[xshift=10.9cm]
            \node[minimum height=1cm, minimum width=1.5cm, fill=red!15, label={\faVolumeUp}] at (\posa,  \posx) (speech-input-1) {};
            \node[data] at (\posa,  \posx) (speech-input) {speech\\embeddings};
            \node[dims] at (\posa, 0.5) (speech-input-dims) {$D_{\aud} \times T$};

            \node[minimum height=1cm, minimum width=1.1cm, fill=orange!20, label={\faImage}] at (\posa,  \posy) (visual-input-1) {};
            \node[data] at (\posa, \posy) (visual-input) {image\\embeddings};
            \node[dims] at (\posa, -1.3) (visual-input-dims) {$D_{\vis} \times K$};

            \node[op] at (\posbc, \posy) (visual-pi) {$\pi_D$};
            \node[op] at (\posbc, \posx) (concat) {$\bowtie_T$};

            \node[minimum height=1cm, minimum width=2.6cm, fill=red!15] at (\posde,  \posx) (speech-visual-1) {};
            \node[minimum height=1cm, minimum width=1.1cm, fill=orange!20] at ($(\posde,  \posx) + (1.0, 0.0)$) (speech-visual-2) {};
            \node[data] at (\posde,  \posx) (output) {output\\embeddings};
            \node[dims] at (\posde, 0.5) (output-dims) {$D_{\aud} \times (T + K)$};

            \draw [->] (speech-input) -- (concat);
            \draw [->] (visual-input) -- (visual-pi);
            \draw [->] (visual-pi) -- (concat);
            \draw      (concat) -- (output);

            \begin{pgfonlayer}{background}
                \path (visual-input-1.south -| speech-input-1.west) + (-0.1, -0.1) node (bottom left) {};
                \path (concat.north -| speech-visual-2.east) + (0.1, 1.0) node (top right) {};
                \filldraw [gray!5] (top right) rectangle (bottom left);
                \path (visual-input-1.south -| speech-visual-2.east) + (-0.4, 0.2) node[blue!50!gray] (bottom right) {\method{seq}}; 
            \end{pgfonlayer}
        \end{scope}

    \end{tikzpicture}
    }
    \caption{%
        The two proposed fusion mechanisms of the audio and visual modalities:
        \method{emb}, fuses along the embedding dimension (left);
        \method{seq}, fuses along the sequence dimension (right).
        Additional operations (dense projection, denoted by $\pi$; repeat operation, denoted by $\rho$) ensure matching dimensions and better adapted embeddings;
        the subscript ($D$ or $T$) indicates the axis along which each transformation is applied (embedding dimension or sequence dimension).
        The symbol ``$\bowtie$'' denotes concatenation.
    }
    \label{fig:fusion}
\end{figure*}
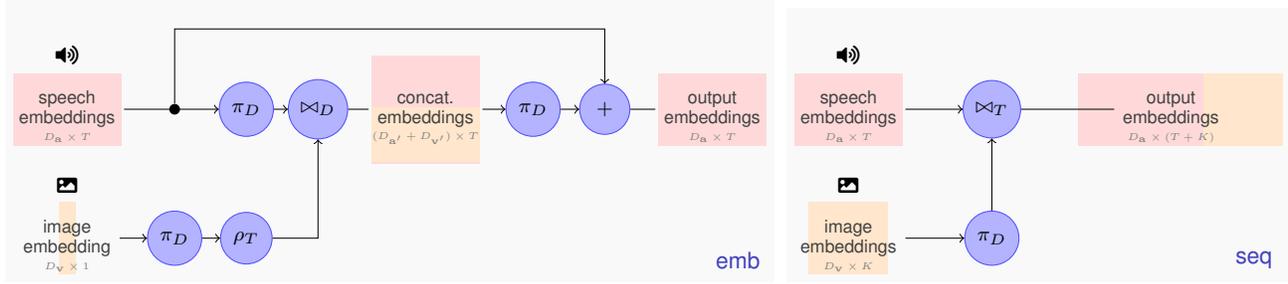

In speech recognition, an audio input $\aud$ is mapped to a transcription $\txt$, usually represented as a sequence of tokens.
In the usual encoder-decoder instantiation the output is obtained by composing the two components: $\txt = \dec(\enc(\aud))$.
In the case of multimodal speech recognition, we assume that we have access to an additional input---the visual channel $\vis$.
The visual information is processed by a separate encoder, $\encv$, and integrated into the network by a fusion function, which we denote by ``$\fuse$'':
\[
    \txt = \dec(\enc(\aud) \fuse \encv(\vis)).
\]
Next, we discuss each of the components:
speech encoder and decoder in \S\ref{subsec:speech-recognition},
visual encoder in \S\ref{subsec:visual-encoder},
fusion in \S\ref{subsec:fusion}.

\subsection{Speech recognition system}
\label{subsec:speech-recognition}

The backbone of the multimodal system the is an end-to-end automatic speech recognition system.
We use a Transformer network, which is based on self-attention modules for the encoder and attention modules in the decoder to pool information from the audio stream.
The network predicts tokens in an autoregressive fashion, by modelling the probability of the next token given the audio and previously predicted tokens, $p(t_k|\txt_{<k}, \aud)$.

\textbf{Transfer learning.}
Instead of starting the training of the multimodal speech recognition system from scratch,
we explore initializing the speech components (encoder and decoder) from a pretrained speech recognition model.
The base system is a pretrained ASR system on a large unimodal dataset, the LibriSpeech corpus \cite{panayotov2015icassp},
whose weigths we transfer and then adapt on the target multimodal dataset via finetuning.

\textbf{Data augmentations.}
We extend the set of speech samples with perturbed versions of the signal in order to make the system more robust and to encourage the decoder to attend to the visual component.
The augmentations are based on SpecAugment \cite{park2019interspeech} and include time warping, frequency masking and time masking.
The same transformations were used for training the base unimodal system on LibriSpeech and we also apply them when training the multimodal system.
Previous approaches \cite{srinivasan2019icml,srinivasan2020icassp,srinivasan2020bemnlp} used temporal masking,
but in their case the removed segments corresponded to words (such as nouns and places).
As such, these methods rely on additional components such as audio-text alignment and part-of-speech tagging,
while our approach is unstructured and, consequently, free of these dependencies.
Moreover, these methods investigated masking with a different goal in mind (not as a data augmentation technique):
to quantify how well the visual component is able to retrieve the masked words at test time.

\subsection{Visual encoder}
\label{subsec:visual-encoder}

The visual encoder summarizes the information present at the input of the visual channel.
We assume an image at input and build upon the popular ResNet architecture \cite{he2016cvpr},
which was also used in previous works on multimodal speech recognition, \eg, \cite{caglayan2019icassp,srinivasan2019icml,srinivasan2020icassp}.
The visual encoder is initialized with the weights of a pretrained model on the ImageNet dataset \cite{russakovsky2015ijcv} and uses intermediate network activations as its encoding.
Depending at which layer we take the activations, we obtain either \ia a single feature vector or \ib a sequence of feature vectors.
Concretely, the activations before the softmax layer (and after the global average pooling layer) yield a single fixed-sized vector,
which encodes global information from the entire image.
If we take the activations from one layer before (that is, before the global average pooling layer), we obtain a 7 $\times$ 7 grid of embeddings, which we sequence as a list of $K = 49$ embeddings.
This second approach encodes more local information, which we hope will allow the model to use more fine-grained characteristics of the image.
On top of the sequence of embeddings we optionally learn layers of gated multilayer perceptrons (gMLP) \cite{liu2021arxiv},
a recently introduced substitute for self-attention layers, which alternates channel-wise with sequence-wise dense layers.
Compared to the attention layer, the gMLP architecture requires less computation and memory, while still maintaining the performance.

\textbf{Relation to prior work.}
Most of prior work uses a single global feature vector to encode the visual information,
some notable exceptions being \cite{paraskevopoulos2020acl} and \cite{srinivasan2020bemnlp}.
The method in \cite{paraskevopoulos2020acl} works on video sequences and extracts a feature vector for each video frame,
while in \cite{srinivasan2020aemnlp} the authors extract ResNet features for $K = 16$ object proposals obtained from a detection network.
Compared to the latter approach, our approach does not require a pretrained detection module and hence is simpler and can be trained with less supervision.

\subsection{Fusion mechanisms}
\label{subsec:fusion}

Our fusion techniques combine the speech and visual embeddings (as produced by each of the two encoders) before feeding them into the decoder.
In the following
we assume that speech embeddings have dimension $D_{\aud} \times T$,
while visual embeddings have dimension $D_{\vis} \times K$ (the second axis, of length $K$, can correspond to a list of boxes in an image or a list of frames in a video).
We experiment with two fusion approaches, either along the embedding dimension (\method{emb}) or along the sequence dimensions (\method{seq});
these two variants are illustrated in Figure \ref{fig:fusion}.
The choice of fusion is also influenced by the visual encoder:
if we represent the visual input with a single feature vector ($K = 1$) it is possible to concatenate along the embedding dimension,
while if we use a list of visual features ($K > 1$) then the concatenation along the sequence dimension is more suitable.

\textbf{Fusing along the embedding dimension} (\method{emb}).
In this case we fuse the speech and visual features along the dimension of the embeddings.
More precisely,
we first project the two inputs to new subspaces (of dimension $D_{\aud'}$ and $D_{\vis'}$, respectively),
replicate the visual embedding $T$ times,
then we concatenate the two representations and, finally, project the output to have dimension $D_{\aud}$.
In this case, the fusion procedure outputs a matrix of the same size as the input speech matrix, $D_{\aud} \times T$.
Retaining the original dimension has a number of advantages:
it allows us to maintain the same decoder size as in the unimodal case (enabling transfer learning and fairer comparisons)
and to use residual connections (from speech to the fused features), which are known to help learning.

\textbf{Fusing along the sequence dimension} (\method{seq}).
When the embeddings of the two input modalities are both sequences, it makes sense to concatenate the visual and speech features along the sequence dimension (temporal for speech and patch-wise for the image).
As the decoder attends along sequential dimension of the input, this operation will become more expensive after the fusion.
However, the \method{seq} fusion has the advantage of being more flexible than the \method{emb} variant,
since the decoder has the option of pooling separately the audio and visual features, without mixing the two.

\textbf{Relation to previous work.}
Many of the previous approaches were based on recurrent networks
and the common ways of incorporating the visual context were
\ia to set the first decoded ``word'' as the visual embedding \cite{sun2016slt,moriya2018slt}, or
\ib to initialize the hidden state of the recurrence with the visual embedding \cite{caglayan2019icassp}.
Another method, encountered especially for adapting acoustic features, was visual adaptive training \cite{miao2016interspeech,palaskar2018icassp},
which amounts to applying a linear transformation parameterized by the visual encoding.
While concatenation of features was previously employed \cite{miao2016interspeech,palaskar2018icassp} it was not used in the context of Transformer architectures.
When the visual information is a sequence,
attention-based methods are a popular choice \cite{srinivasan2020aemnlp,paraskevopoulos2020acl,ghorbani2021slt}.
All these methods pool independently across the audio and visual streams, whereas in our case the \method{seq} method pools over both of them simultaneously.
The methods in \cite{srinivasan2020aemnlp,ghorbani2021slt} attend to the visual sequence based on the previously decoded word (as we do),
while \cite{paraskevopoulos2020acl} pools based on the audio. 
The main distinction between \cite{ghorbani2021slt} and \cite{srinivasan2020aemnlp}
is that the former simply concatenates the two pooled representations,
while the latter predicts which of the two modalities (visual or audio) should be preferred through a second, hierarchical attention layer.

\section{Experimental setup}

In this section we present the experimental setup, including 
the multimodal datasets (\S\ref{subsec:datasets})
and additional implementation details (\S\ref{subsec:implementation-details}).

\subsection{Datasets}
\label{subsec:datasets}

We carry out the evaluation on three datasets that contain the three desired modalities (audio, visual, language).

\textbf{Flickr8K} \cite{hodosh2013jair,harwath2015asru}
consists of 8K images, each described by five captions.
The original dataset \cite{hodosh2013jair} contained only the visual and language modalities,
and it was later extended with audio recordings of the read captions by \cite{harwath2015asru}.

\textbf{How2} \cite{sanabria2018nips}
contains instructional videos downloaded from YouTube
and comes with additional shot information and transcriptions.
We use the 300h variant, which totals around 13.5K videos (190K shots).
The dataset consists of pre-extracted audio and visual features,
but, in order to use pretrained models, we had to use the original videos;
the raw data was kindly provided by the authors, upon request.

\textbf{Localized Narratives} \cite{pont2020eccv}
is a recently introduced dataset, that extends four popular image datasets (Flickr30K \cite{young2014tacl}, COCO \cite{lin2014eccv}, ADE20K \cite{zhou2019ijcv}, Open Images \cite{kuznetsova2020ijcv})
with new captions, audio recordings and mouse traces (which locate the spoken words in the image).
Compared to the original datasets,
the captions are richer and the audio component is challenging due to noisy recording conditions and accented speech.
We use this dataset to carry out an ablation study.
In order to be able to perform such extensive studies we
\ia use only the Flickr30K part,
\ib segment the audio into sentences (based on the provided transcripts),
\ic remove utterances longer than 15 seconds,
\id subsample half of the utterances.
This procedure is applied on all three splits (training, validation, testing) and
yields around 32K, 1K, 1K samples, respectively.


\subsection{Implementation details}
\label{subsec:implementation-details}

Our implementation is based on the ESPnet framework \cite{watanabe2018interspeech} and
the code is available online.%
\footnote{\small\url{https://github.com/danoneata/espnet/tree/multimodal-asr/egs2/vorbis}}

The speech recognition component is a Transformer architecture and is pretrained on the LibriSpeech dataset \cite{panayotov2015icassp};
this model achieves 2.6\% word error rate on the LibriSpeech \texttt{clean} test set and 6.0\% on the LibriSpeech \texttt{other} test set.
The audio encoder has 18 blocks and outputs embeddings with size 512,
while the decoder has six layers and outputs tokens from a vocabulary with 5000 elements.
The vocabulary was obtained by subword segmentation using an unigram language model \cite{kudo2018acl};
we don't use an external language model.
For finetuning, we train for 50 epochs for the smaller datasets (Flickr8K and Localized Narratives) and 30 epochs for the larger dataset (How2).
For optimization, the learning rate is warmed up linearly from $3.2 \times 10^{-8}$ to $8 \times 10^{-4}$ over $25$K batches,
after which it is decreased as a function of $1 / s^2$ in the step number $s$.
At test time, we predict using a model whose weights are set to the average of the ten best checkpoints encountered during training;
this technique is applied to all of our reported results and it yields small but consistent improvements (around 5\% relative improvement) over predicting with only the best model.

The visual encoder is a ResNet architecture with either 18 or 50 layers, pretrained on the ImageNet dataset \cite{russakovsky2015ijcv},
yielding 512 or 2048-dimensional embeddings, respectively.
The input image is rescaled to 224 $\times$ 224 pixels and standardized using the ImageNet statistics.
We perform image data augmentation by random horizontal flipping.
The How2 dataset contains video, but since our visual embedding works on images, we use only the middle frame.
As the videos are shots and hence stable in terms of viewpoint change,
we expect a single frame to encode enough information.

For the \method{emb} fusion variant we project both the audio and visual embeddings to a 128-dimensional space.
For the \method{seq} variant we use two gMLP layers on top of the visual embeddings,
as this choice gave slightly better results than the alternatives of using zero or one layer.

\section{Experiments}

\begin{table}
    \centering
    \small
    \begin{tabular}{lccccc}
        \toprule
        method & visual & fuse & Flickr8K & How2 & Loc. Nar. \\
        \midrule
        \multirow{2}{*}{\cite{sun2016slt}}              &          &  & 14.8 & ---  & --- \\
                                                        & \faCheck &  & 13.8 & ---  & --- \\
        \multirow{2}{*}{\cite{srinivasan2020aemnlp}}    &          &  & 13.6 & ---  & --- \\
                                                        & \faCheck &  & 14.1 & ---  & --- \\
        \multirow{2}{*}{\cite{paraskevopoulos2020acl}}  &          &  & ---  & 19.2 & --- \\
                                                        & \faCheck &  & ---  & 18.4 & --- \\
        \multirow{2}{*}{\cite{ghorbani2021slt}}         &          &  & ---  & 17.7 & --- \\
                                                        & \faCheck &  & ---  & 17.2 & --- \\
        \midrule
        \methodfs{pretrain} &          &                & 11.1    & 26.9     & 49.3 \\
        \methodfs{finetune} &          &                & \bf 3.8 & 11.8     & 4.3 \\
        \methodfs{finetune} & \faCheck & \methodfs{emb} & 4.3     & 11.1     & \bf 3.9 \\
        \methodfs{finetune} & \faCheck & \methodfs{seq} & 4.7     & \bf 10.8 & 4.0 \\
        \bottomrule
    \end{tabular}
    \caption{%
        Comparison to state-of-the-art approaches on the test sets of three multimodal datasets (Flickr8K, How2 and Localized Narratives) in terms of word error rate (lower values are better).
        Visual indicates those variants that use the visual channel as input in addition to the speech.
    }
    \label{tab:sota}
\end{table}

This section presents the empirical evaluation of the proposed methodology.
First, in \S \ref{subsec:main-results} we compare our best unimodal and multimodal systems to baseline and state-of-the-art approaches.
Second, in \S \ref{subsec:ablation} we present an ablation study over the main individual contributions: transfer learning and data augmentations.

\subsection{Main results}
\label{subsec:main-results}

Table \ref{tab:sota} presents speech recognition performance for four of our systems:
two unimodal variants (a \method{pretrained} ASR, used as a baseline, and its \method{finetuned} counterpart based on adapting the \method{pretrained} method on each dataset) and
two multimodal variants (both trained by finetuning all components, but differing in the fusion techniques, \method{emb} or \method{seq}, as described in \S \ref{subsec:fusion}).
Both multimodal methods use the SpecAugment data augmentation and a ResNet with 50 layers as the visual encoder.
We compare our approaches to state-of-the-art methods.
Previous studies evaluate usually on a single dataset, for example, Flickr8K \cite{sun2016slt,srinivasan2020aemnlp} or How2 \cite{paraskevopoulos2020acl,ghorbani2021slt},
while we report performance on three datasets:
the two aforementioned ones and Localized Narratives (on which we are the first to report multimodal speech recognition performance).

We observe that the \method{pretrained} method already improves over previous work on Flickr8K, although its results are poorer on How2 and Localized Narratives due to data mismatch.
However, by \method{finetuning}, the speech-only unimodal system significantly outperforms the current state-of-the-art,
yielding relative improvements of 72\% and 31\% on Flickr8K and How2, respectively.
The results for the multimodal systems, which include the visual information, are better than the unimodal results in the case of How2 and Localized Narratives dataset;
for Flickr8K it is difficult to improve presumably because it is a clean dataset for which the ASR already works well and many of its errors are not visually grounded.
Among the two fusing methods the results are mixed, the fusion along the embedding dimension, \method{emb}, being the better method on two of the three datasets.




\subsection{Ablation studies}
\label{subsec:ablation}

In this subsection we carry two main ablation studies to better understand the impact of data augmentation and the importance of transferring representations.

\begin{table}
    \centering
    \small
    \begin{tabular}{lcccccc}
        \toprule
        visual    & fuse & aug.     &  & Flickr8K & How2     & Loc. Nar. \\
        \midrule
                   & ---            &          &  & \bf 3.8 & 11.8     & \bf 4.3 \\
                   & ---            & \faCheck &  & 4.2     & \bf 11.2 & 4.5 \\
        \midrule
        \faCheck   & \methodfs{emb} &          &  & 4.8     & 11.8     & 4.1 \\
        \faCheck   & \methodfs{emb} & \faCheck &  & \bf 4.3 & \bf 11.1 & \bf 3.9 \\
        \midrule
        \faCheck   & \methodfs{seq} &          &  & \bf 4.0 & 11.8     & 4.2 \\
        \faCheck   & \methodfs{seq} & \faCheck &  & 4.7     & \bf 10.8 & \bf 4.0 \\
        \bottomrule
    \end{tabular}
    \caption{%
        Evaluation of the impact of audio augmentations (aug.) on the test sets of the three multimodal datasets in terms of word error rate.
        All models are \methodfs{finetuned} and the multimodal variants use the ResNet50 as visual encoder.
    }
    \label{tab:specaug}
\end{table}

\textbf{Data augmentations.}
We evaluate the impact of the SpecAugment data augmentation technique in three scenarios:
for the unimodal system and for the two multimodal variants using the two feature fusion techniques (\method{emb} and \method{seq}).
For all cases, we perform finetuning of all components and for the visual-based systems we use the 50-layer ResNet.

Table \ref{tab:specaug} reveals that speech data augmentation is important for the multimodal systems,
yielding improvements in five out of the six cases.
These results suggest that perturbing the audio signal might help the multimodal models rely more on the visual encoder and eventually produce better results.
Surprisingly, the unimodal variants have seen little benefit from data augmentation, with the exception of the results on the How2 dataset.




\begin{table*}
    \centering
    \newcommand{\ii}[1]{\footnotesize{\color{gray}#1}}
    \small
    \begin{tabular}{ccclccclcclrr}
        \toprule
        & \multicolumn{2}{c}{audio encoder} & & \multicolumn{3}{c}{visual encoder} & & \multicolumn{2}{c}{decoder} & & \multirowcell{2}{WER (\%)} & \multirowcell{2}{num. trainable\\params ($\times10^6$)} \\
        \cmidrule(lr){2-3} \cmidrule(lr){5-7} \cmidrule(lr){9-10}
                 & init      & train     &  & init     & train     & network  &  & init      & train     &  \\
        \midrule
        \ii{1}      & \faShare & \faLock     &  & ---      & ---         & ---      &  & \faShare & \faLock     &  & 49.3 & 0 \\ 
        \ii{2}      & \faDice  & \faLockOpen &  & ---      & ---         & ---      &  & \faDice  & \faLockOpen &  & 22.5 & 99.4 \\ 
        \ii{3}      & \faShare & \faLock     &  & ---      & ---         & ---      &  & \faDice  & \faLockOpen &  & 9.1  & 32.9 \\ 
        \ii{4}      & \faShare & \faLock     &  & ---      & ---         & ---      &  & \faShare & \faLockOpen &  & 6.3  & 32.9 \\ 
        \ii{5}      & \faShare & \faLockOpen &  & ---      & ---         & ---      &  & \faShare & \faLockOpen &  & 4.3  & 99.4 \\ 
        \midrule
        \ii{6}      & \faShare & \faLock     &  & \faShare & \faLock     & ResNet18 &  & \faShare & \faLockOpen &  & 5.8  & 33.2 \\ 
        \ii{7}      & \faShare & \faLock     &  & \faShare & \faLockOpen & ResNet18 &  & \faShare & \faLockOpen &  & 5.5  & 44.4 \\ 
        \ii{8}      & \faShare & \faLockOpen &  & \faShare & \faLock     & ResNet18 &  & \faShare & \faLockOpen &  & 4.3  & 99.6 \\ 
        \ii{9}      & \faShare & \faLockOpen &  & \faShare & \faLockOpen & ResNet18 &  & \faShare & \faLockOpen &  & 4.2  & 110.8 \\ 
        \midrule
        \ii{10}     & \faShare & \faLock     &  & \faShare & \faLock     & ResNet50 &  & \faShare & \faLockOpen &  & 5.9  & 33.4 \\ 
        \ii{11}     & \faShare & \faLock     &  & \faShare & \faLockOpen & ResNet50 &  & \faShare & \faLockOpen &  & 5.6  & 56.9 \\ 
        \ii{12}     & \faShare & \faLockOpen &  & \faShare & \faLock     & ResNet50 &  & \faShare & \faLockOpen &  & 4.1  & 99.8 \\ 
        \ii{13}     & \faShare & \faLockOpen &  & \faShare & \faLockOpen & ResNet50 &  & \faShare & \faLockOpen &  & 4.1  & 123.3 \\ 
        \bottomrule
    \end{tabular}
    \caption{%
        Transferring representations---evaluation on the test set of the Localized Narratives dataset.
        For each of the three components of the model (audio encoder, visual encoder, decoder), we indicate
        how the model's weights are initialized (either random \faDice\xspace or shared from a pretrained model \faShare) and trained (either fixed \faLock\xspace or finetuned \faLockOpen).
        For each setting we report the word error rate (WER) and the number of trainable parameters.
        The visual information is fused with the \methodfs{emb} method. 
        For these experiments, we did not use audio augmentation.
    }
    \label{tab:transfer-representations}
\end{table*}

\textbf{Transferring representations.}
We conduct an extensive ablation study over the impact of
initialization (random or pretrained)
and training procedure (fixed or finetuned weights)
for each of the components of the model (audio encoder, visual encoder, decoder).
These experiments are carried on the Localized Narratives dataset
and for the multimodal setting we use only the \method{emb} fusing method.
Additionally, we investigate the impact of visual encoder's capacity by varying the number of layers in the ResNet architecture: 18 or 50.
The results are presented in Table \ref{tab:transfer-representations}.

Rows 1--5 show the results for the unimodal system, corresponding to a standard, speech-only ASR.
We observe that the \method{pretrained} variant, without any finetuning (row 1) is underperforming,
most likely, due to the large mismatch in both terms of audio (speech is noisy and accented) and language data.
On the other hand, ignoring the availability of pretrained representations (row 2) is also not ideal:
training the network from scratch, as is customary done in previous works, produces better, but still unsatisfactory transcriptions.
Rows 3 and 4 show the results for the case when the encoder is fixed and the decoder is trained: either from scratch (row 3) or by finetuning the pretrained weights (row 4).
Since the decoder of an end-to-end ASR model plays also the role of a language model,
this procedure is akin to language adaptation and results in significant boosts in performance for both variants.
Finally, finetuning both components (row 5) yields the best results, with a relative improvement of around 30\%.

%


Rows 6--13 present the results for the multimodal systems, using a visual encoder with 18 (rows 6--9) or 50 layers (rows 10--13).
For this set of experiments, we use only with the finetuning approach, as the results of the unimodal system have showed that this technique is superior.
We also always adapt the decoder because the speech-vision fusion affects the distribution of features.
Note that the fused features are projected to the same embedding dimension as the speech features, which enables sharing the pretrained decoder weights.
The projection layers in the fusion layer are always trainable.


We first note that including visual information improves over the single-stream system in all scenarios:
either if we keep the encoders fixed (rows 6 and 10 vs row 4) or if we finetune the encoders (rows 9 and 13 vs row 5).
Second, we observe that we obtain better results as we allow for more components to be finetuned,
with the last column indicating a correlation between the number of trainable parameters and the performance.
The best results are achieved when finetuning all components (rows 9 and 13).
Finally, increasing the capacity of the visual encoder yields similar results.
We do see slight improvements for the cases when we finetune the speech encoder (row 12 vs row 8; row 13 vs row 9),
potentially suggesting the coupling between the two modalities needs to be accounted also by the encoders and not only the decoders.

\subsection{Qualitative results}
\label{subsec:qualitative-results}

\begin{table*}
    \centering
    \footnotesize
    \def\ww{3.5cm}
    \def\hh{2.5cm}
    \newcommand\mistake[1]{{\color{red}\textit{#1}}}
    \begin{tabularx}{\linewidth}{YYY}
        \includegraphics[width=\ww, height=\hh]{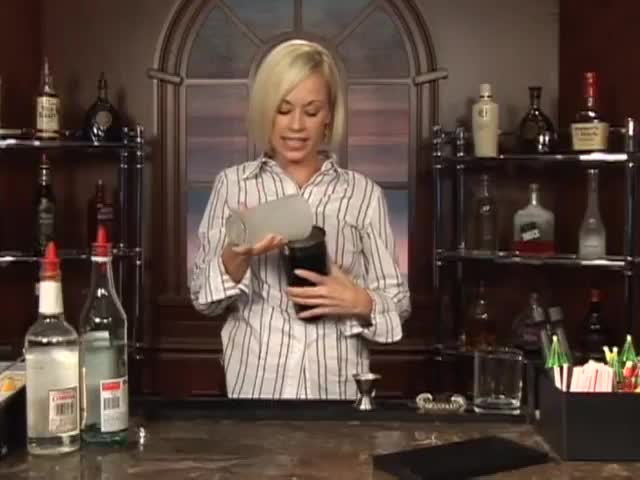}                                 & \includegraphics[width=\ww, height=\hh]{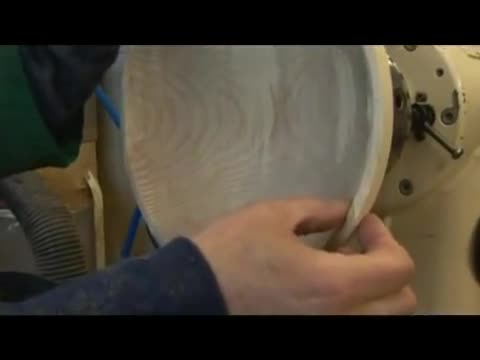}                                            & \includegraphics[width=\ww, height=\hh]{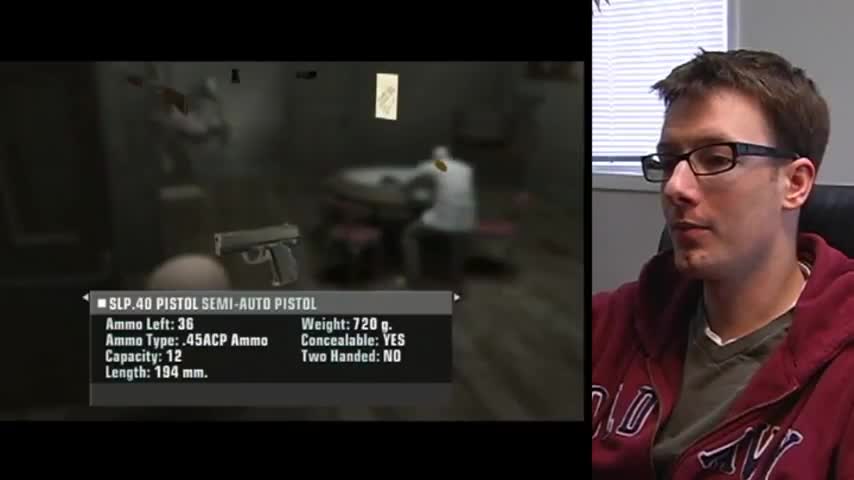} \\
        \textsf{r} · mix it up really good because that egg white is thick it's really thick             & \textsf{r} · and spalting is nothing more than the natural decay process that wood goes through            & \textsf{r} · so we can either take the fiber wire or use the shotgun or use the pistol that we picked up \\
        \textsf{u} · mix it up really good because that \mistake{eight} white is thick it's really thick & \textsf{u} · and spalting is nothing more than the natural decay process that \mistake{would} goes through & \textsf{u} · so we can either take the fiber wire or use the \mistake{shock on} or use the pistol that we picked up \\
        \textsf{m} · mix it up really good because that egg white is thick it's really thick             & \textsf{m} · and spalting is nothing more than the natural decay process that wood goes through            & \textsf{m} · so we can either take the fiber wire or use the shotgun or use the pistol that we picked up \\
        \midrule
        \includegraphics[width=\ww, height=\hh]{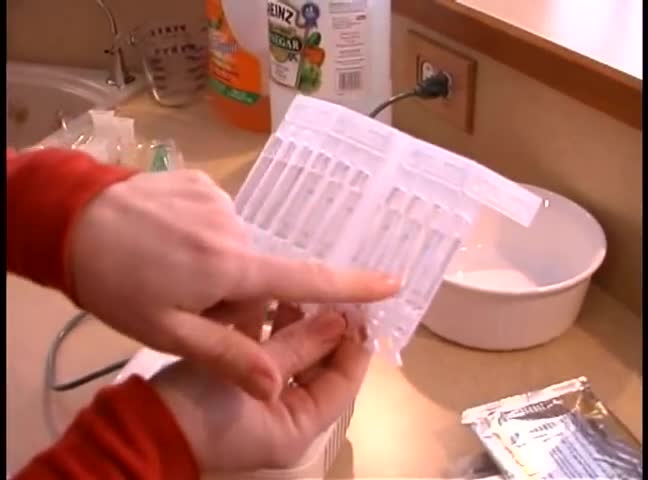}                                  & \includegraphics[width=\ww, height=\hh]{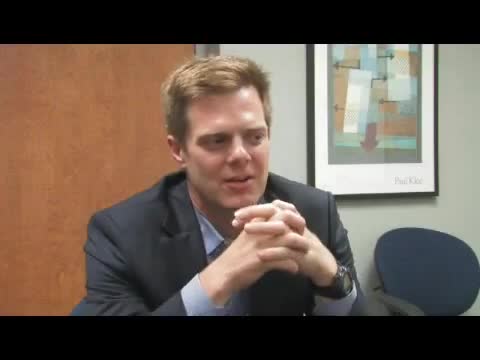}                                           & \includegraphics[width=\ww, height=\hh]{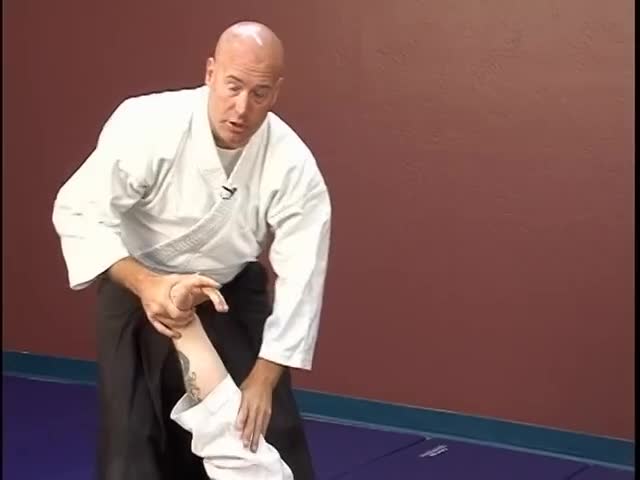} \\
        \textsf{r} · so each vial here is actually one use                                               & \textsf{r} · that's just a long road                                                                       & \textsf{r} · never leave your person on their back \\
        \textsf{u} · so each \mistake{vile} here is actually one use                                     & \textsf{u} · that is just a long \mistake{row}                                                             & \textsf{u} · never leave your person \mistake{on your} on their back \\
        \textsf{m} · so each vial here is actually one use                                               & \textsf{m} · that's just a long road                                                                       & \textsf{m} · never leave your person on their back \\
        \midrule
        \includegraphics[width=\ww, height=\hh]{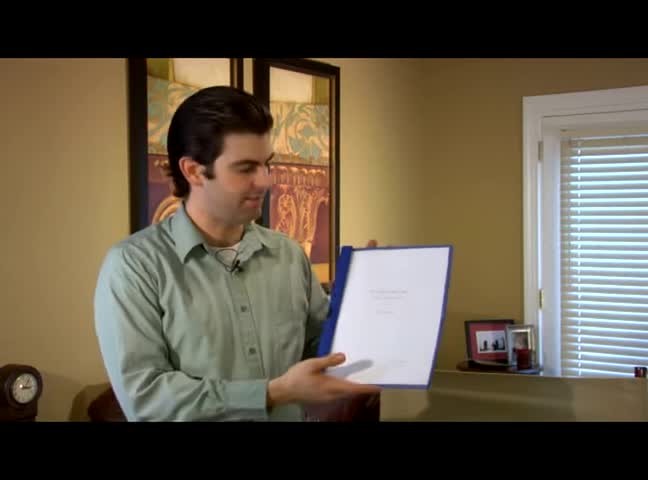}                                  & \includegraphics[width=\ww, height=\hh]{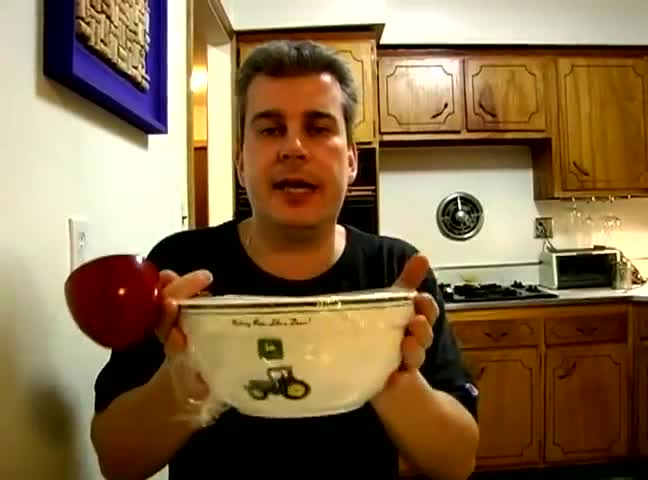}                                            & \includegraphics[width=\ww, height=\hh]{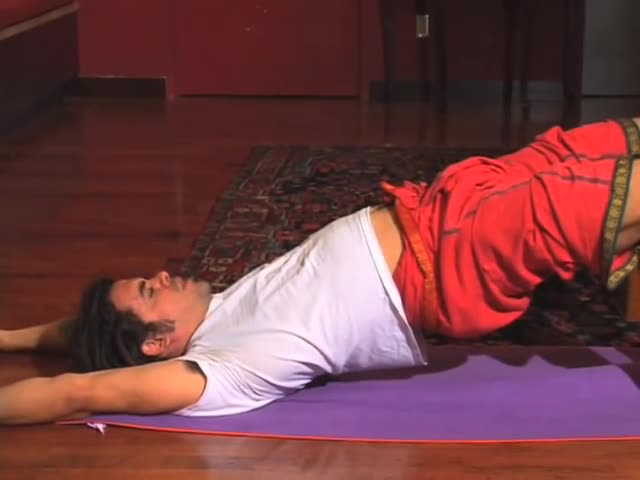} \\
        \textsf{r} · here's an example of a nicely bound script                                          & \textsf{r} · i just took our salad out of the refrigerator                                                 & \textsf{r} · five more keep breathing deep expanding \\
        \textsf{u} · here's an example of a nicely bound script                                          & \textsf{u} · i just took our salad out of the refrigerator                                                 & \textsf{u} · five more keep breathing deep expanding \\
        \textsf{m} · here's an example of a nicely \mistake{balance} script                              & \textsf{m} · \mistake{now} i just took our salad out of the refrigerator                                   & \textsf{m} · five more \mistake{key breathe} than deep expanding \\
    \end{tabularx}
    \caption{%
        Qualitative results on the How2 dataset.
        For each example, we show the central frame of the video shot, the reference text (\textsf{r}), the transcriptions using the unimodal model (\textsf{u}) and the multimodal one (\textsf{m}).
        The mistakes are shown in red and italics.
        The multimodal model is using the \method{finetune} setting, the \method{emb} fusing method and the ResNet50 visual encoder.
    }
    \label{tab:qualitative-results}
\end{table*}

In this subsection we present a series of qualitative examples to contrast the output of the unimodal system with the one of the multimodal system.
The results are provided on the How2 dataset using models trained in the \method{finetune} regime and using data augmentation;
for the multimodal model we used the \method{emb} fusing method and the ResNet50 visual encoder.
Table \ref{tab:qualitative-results} shows nine samples where the two systems differ in their predictions, but at least one agrees with the groundtruth transcriptions:
for the first two rows the multimodal system is correct, while in the last row the unimodal system is correct.
The first four examples are cases where the multimodal system correctly chooses between phonetically similar words,
for example, it corrects ``eight'' to ``egg'', ``would'' to ``wood'', ``shock on'' to ``shotgun'', ``vile'' to ``vial''.
While these examples suggest that the visual context helps the transcription,
the rest of the samples are more difficult to interpret.



\subsection{Discussion and limitations}
\label{subsec:limitations}


We discuss two potential limitations of our work.
These shortcomings stem from the nature of the problem, which involves a modality (the audio input) that dominates the other (the visual input).

\textbf{Limited potential improvements.}
We have seen that the results usually improve under the multimodal setting,
but a possible criticism of our work could be that the gains are not significant.
However, our results are in line with prior art (see Table \ref{tab:sota}), which works in a more favourable setup as their multimodal variants have more scope for improvement.
Moreover, we believe that it is difficult to obtain considerably larger improvements in the current setting.
While accurately quantifying the upper bound of the multimodal performance is a difficult task,
we manually inspect the most common mistakes that occur in the unimodal setting
(we consider the How2 dataset and the unimodal method trained with the \method{finetune} option and using data augmentation).
We observe that many of the most frequent mistakes involve short words that correspond to prepositions, pronouns, indeterminates
(such as ``and'', ``to'', ``you'', ``I'', ``the'', ``a'', ``that''), which are either erroneously inserted or missed in the automated transcription.
These types of mistakes are difficult to correct by the visual modality since they don't have visual grounding.
In the tail of the distribution we find mistakes that have visual context,
such as the pairs presented in the qualitative results or substitution mistakes such as ``cymbal'' $\to$ ``symbol'', ``sprite'' $\to$ ``sprayed'', ``bow'' $\to$ ``boat'' or ``both''.
However, this type of errors are in minority, indicating that substantial improvements are difficult to achieve.

\textbf{On the r\^ole of the visual modality.}
The qualitative results indicate that there are still cases for which it remains unclear \textit{how} the multimodal system uses the visual component.
Previous work in the context of both multimodal speech recognition \cite{srinivasan2019icml} and multimodal machine translation \cite{wu2021acl} have observed that the visual channel helps in unexpected ways.
In particular, Wu \etal \cite{wu2021acl} suggest that the visual branch plays the role of a regularizer and is not necessarily injecting useful information into the system.
In our case, while we believe that the proposed speech data augmentation may alleviate the issue to some extent, its effect may remain still insufficient.
We conjecture that the problem lies in the loose coupling of the input modalities.
A possible solution for a more pervasive fusion would be pretraining a self-supervised audio-visual model on large quantities of data.
Such audio-visual systems are becoming common place, but they were not usually applied in the multimodal speech recognition setting.
The closest works in this direction are the ones of
Hsu \etal \cite{hsu2019interspeech}, which uses the pretrained audio-visual representations for unimodal speech recognition, and
Rouditchenko \etal \cite{rouditchenko2021interspeech}, which performs multimodal text retrieval.

\section{Conclusions}
\label{sec:conclusions}

In this paper, we extend and build upon state-of-the-art approaches for multimodal speech recognition.
We employ a Transformer ASR architecture as a baseline system in which we inject visual information through a ResNet image encoder.
In contrast to the previous methods, we leverage pretrained representations for both the speech and visual channels, and we further explore two fusion techniques for the two input channels.
Our approach leads to substantial improvements over the state of the art on two standard multimodal datasets: Flickr8K and How2.
The ablation studies provide important insights
on the role of using speech augmentation before training the multimodal network and
the individual contribution of finetuning the various components of the system.
While we find it remarkable that the multimodal setting still improves over a strong unimodal baseline,
the qualitative examples still leave open the question of how system is using the visual information.

\textbf{Future work.}
To better understand the complicated interactions that arise in a multimodal speech recognition network,
a future work direction could investigate which parts of the inputs (audio, image, previously predicted tokens) contribute more to the output;
modern tools for explainable machine learning \cite{kokhlikyan2020arxiv,samek2021ieee,joshi2021access} are a possible avenue for this enterprise.
Another future direction could involve coupling the two encoders through a more pervasive fusion by leveraging self-supervised audio-visual models that are pretrained on large-scale datasets.

\textbf{Broader impact.}
The goal of this research work is making speech recognition more accurate when additional visual information is available.
Speech recognition is a technology that enables many impactful applications, such as automatic caption generation or keyword search in large collections of audio data.
While as any machine learning approach it can be subject to biases in the data or malicious use,
we do not foresee any specific societal harm that could stem from the current line of work.


\textbf{Acknowledgements.}
We thank the anonymous reviewers and Desmond Elliott for useful suggestions.
This work was supported in part by a grant of the Romanian Ministry of Education and Research, CNCS - UEFISCDI, project number PN-III-P1-1.1-PD-2019-0918, within PNCDI III.

{\small
\bibliographystyle{ieee_fullname}
\bibliography{ref}

\begin{thebibliography}{10}\itemsep=-1pt

\bibitem{afouras2018pami}
Triantafyllos Afouras, Joon~Son Chung, Andrew Senior, Oriol Vinyals, and Andrew
  Zisserman.
\newblock Deep audio-visual speech recognition.
\newblock {\em IEEE Transactions on Pattern Analysis and Machine Intelligence},
  2018.

\bibitem{caglayan2019icassp}
Ozan Caglayan, Ramon Sanabria, Shruti Palaskar, Loic Barraul, and Florian
  Metze.
\newblock Multimodal grounding for sequence-to-sequence speech recognition.
\newblock In {\em IEEE International Conference on Acoustics, Speech and Signal
  Processing}, pages 8648--8652, 2019.

\bibitem{fleischman2008acl}
Michael Fleischman and Deb Roy.
\newblock Grounded language modeling for automatic speech recognition of sports
  video.
\newblock In {\em Association for Computational Linguistics}, pages 121--129,
  2008.

\bibitem{ghorbani2021slt}
Shahram Ghorbani, Yashesh Gaur, Yu Shi, and Jinyu Li.
\newblock Listen, look and deliberate: Visual context-aware speech recognition
  using pre-trained text-video representations.
\newblock In {\em IEEE Spoken Language Technology Workshop}, pages 621--628,
  2021.

\bibitem{gupta2017icassp}
Abhinav Gupta, Yajie Miao, Leonardo Neves, and Florian Metze.
\newblock Visual features for context-aware speech recognition.
\newblock In {\em IEEE International Conference on Acoustics, Speech and Signal
  Processing}, pages 5020--5024, 2017.

\bibitem{harwath2015asru}
David Harwath and James Glass.
\newblock Deep multimodal semantic embeddings for speech and images.
\newblock In {\em Workshop on Automatic Speech Recognition and Understanding},
  pages 237--244, 2015.

\bibitem{harwath2019iclr}
David Harwath, Wei-Ning Hsu, and James Glass.
\newblock Learning hierarchical discrete linguistic units from
  visually-grounded speech.
\newblock In {\em International Conference on Learning Representations}, 2019.

\bibitem{harwath2018eccv}
David Harwath, Adri{\`a} Recasens, D{\'\i}dac Sur{\'\i}s, Galen Chuang, Antonio
  Torralba, and James Glass.
\newblock Jointly discovering visual objects and spoken words from raw sensory
  input.
\newblock In {\em European Conference on Computer Vision}, pages 649--665,
  2018.

\bibitem{harwath2016nips}
David Harwath, Antonio Torralba, and James Glass.
\newblock Unsupervised learning of spoken language with visual context.
\newblock In {\em Advances in Neural Information Processing Systems}, pages
  1858--1866, 2016.

\bibitem{he2016cvpr}
Kaiming He, Xiangyu Zhang, Shaoqing Ren, and Jian Sun.
\newblock Deep residual learning for image recognition.
\newblock In {\em IEEE Conference on Computer Vision and Pattern Recognition},
  pages 770--778, 2016.

\bibitem{hodosh2013jair}
Micah Hodosh, Peter Young, and Julia Hockenmaier.
\newblock Framing image description as a ranking task: Data, models and
  evaluation metrics.
\newblock {\em Journal of Artificial Intelligence Research}, 47:853--899, 2013.

\bibitem{hsu2019interspeech}
Wei-Ning Hsu, David Harwath, and James Glass.
\newblock Transfer learning from audio-visual grounding to speech recognition.
\newblock In {\em Interspeech}, pages 3242--3246, 2019.

\bibitem{joshi2021access}
Gargi Joshi, Rahee Walambe, and Ketan Kotecha.
\newblock A review on explainability in multimodal deep neural nets.
\newblock {\em IEEE Access}, 2021.

\bibitem{kamper2019taslp}
Herman Kamper, Gregory Shakhnarovich, and Karen Livescu.
\newblock Semantic speech retrieval with a visually grounded model of
  untranscribed speech.
\newblock {\em Transactions on Audio, Speech and Language Processing},
  27(1):89--98, 2019.

\bibitem{kokhlikyan2020arxiv}
Narine Kokhlikyan, Vivek Miglani, Miguel Martin, Edward Wang, Bilal Alsallakh,
  Jonathan Reynolds, Alexander Melnikov, Natalia Kliushkina, Carlos Araya, Siqi
  Yan, et~al.
\newblock Captum: A unified and generic model interpretability library for
  {PyTorch}.
\newblock {\em arXiv preprint arXiv:2009.07896}, 2020.

\bibitem{kudo2018acl}
Taku Kudo.
\newblock Subword regularization: Improving neural network translation models
  with multiple subword candidates.
\newblock In {\em Association for Computational Linguistics}, pages 66--75,
  2018.

\bibitem{kuznetsova2020ijcv}
Alina Kuznetsova, Hassan Rom, Neil Alldrin, Jasper Uijlings, Ivan Krasin, Jordi
  Pont-Tuset, Shahab Kamali, Stefan Popov, Matteo Malloci, Alexander
  Kolesnikov, et~al.
\newblock The {O}pen {I}mages dataset v4.
\newblock {\em International Journal of Computer Vision}, pages 1--26, 2020.

\bibitem{lin2014eccv}
Tsung-Yi Lin, Michael Maire, Serge Belongie, James Hays, Pietro Perona, Deva
  Ramanan, Piotr Doll{\'a}r, and C~Lawrence Zitnick.
\newblock Microsoft {COCO}: Common objects in context.
\newblock In {\em European conference on computer vision}, pages 740--755.
  Springer, 2014.

\bibitem{liu2021arxiv}
Hanxiao Liu, Zihang Dai, David~R So, and Quoc~V Le.
\newblock Pay attention to {MLP}s.
\newblock {\em arXiv preprint arXiv:2105.08050}, 2021.

\bibitem{ma2021icassp}
Pingchuan Ma, Stavros Petridis, and Maja Pantic.
\newblock End-to-end audio-visual speech recognition with conformers.
\newblock In {\em IEEE International Conference on Acoustics, Speech and Signal
  Processing}, pages 7613--7617, 2021.

\bibitem{miao2016interspeech}
Yajie Miao and Florian Metze.
\newblock Open-domain audio-visual speech recognition: A deep learning
  approach.
\newblock In {\em Interspeech}, pages 3414--3418, 2016.

\bibitem{moriya2018slt}
Yasufumi Moriya and Gareth~JF Jones.
\newblock {LSTM} language model adaptation with images and titles for
  multimedia automatic speech recognition.
\newblock In {\em IEEE Spoken Language Technology Workshop}, pages 219--226,
  2018.

\bibitem{moriya2019icassp}
Yasufumi Moriya and Gareth~JF Jones.
\newblock Multimodal speaker adaptation of acoustic model and language model
  for {ASR} using speaker face embedding.
\newblock In {\em IEEE International Conference on Acoustics, Speech and Signal
  Processing}, pages 8643--8647, 2019.

\bibitem{mroueh2015icassp}
Youssef Mroueh, Etienne Marcheret, and Vaibhava Goel.
\newblock Deep multimodal learning for audio-visual speech recognition.
\newblock In {\em IEEE International Conference on Acoustics, Speech and Signal
  Processing}, pages 2130--2134, 2015.

\bibitem{mukherjee2003eurospeech}
Niloy Mukherjee and Deb Roy.
\newblock A visual context-aware multimodal system for spoken language
  processing.
\newblock In {\em European Conference on Speech Communication and Technology},
  2003.

\bibitem{oneata2021cee}
Dan Oneață and Horia Cucu.
\newblock Multimodal speech recognition for unmanned aerial vehicles.
\newblock {\em Computers \& Electrical Engineering}, 90:106943, 2021.

\bibitem{palaskar2018icassp}
Shruti Palaskar, Ramon Sanabria, and Florian Metze.
\newblock End-to-end multimodal speech recognition.
\newblock In {\em IEEE International Conference on Acoustics, Speech and Signal
  Processing}, pages 5774--5778, 2018.

\bibitem{panayotov2015icassp}
Vassil Panayotov, Guoguo Chen, Daniel Povey, and Sanjeev Khudanpur.
\newblock Librispeech: An {ASR} corpus based on public domain audio books.
\newblock In {\em IEEE International Conference on Acoustics, Speech and Signal
  Processing}, pages 5206--5210, April 2015.

\bibitem{paraskevopoulos2020acl}
Georgios Paraskevopoulos, Srinivas Parthasarathy, Aparna Khare, and Shiva
  Sundaram.
\newblock Multiresolution and multimodal speech recognition with transformers.
\newblock In {\em Association for Computational Linguistics}, 2020.

\bibitem{park2019interspeech}
Daniel~S Park, William Chan, Yu Zhang, Chung-Cheng Chiu, Barret Zoph, Ekin~D
  Cubuk, and Quoc~V Le.
\newblock Spec{A}ugment: A simple data augmentation method for automatic speech
  recognition.
\newblock In {\em Interspeech}, pages 2613--2617, 2019.

\bibitem{petridis2018arxiv}
Stavros Petridis, Themos Stafylakis, Pingchuan Ma, Feipeng Cai, Georgios
  Tzimiropoulos, and Maja Pantic.
\newblock End-to-end audiovisual speech recognition.
\newblock {\em CoRR}, abs/1802.06424, 2018.

\bibitem{pont2020eccv}
Jordi Pont-Tuset, Jasper Uijlings, Soravit Changpinyo, Radu Soricut, and
  Vittorio Ferrari.
\newblock Connecting vision and language with localized narratives.
\newblock In {\em European Conference on Computer Vision}, pages 647--664,
  2020.

\bibitem{rouditchenko2021interspeech}
Andrew Rouditchenko, Angie Boggust, David Harwath, Dhiraj Joshi, Samuel Thomas,
  Kartik Audhkhasi, Rogerio Feris, Brian Kingsbury, Michael Picheny, Antonio
  Torralba, et~al.
\newblock {AVL}net: Learning audio-visual language representations from
  instructional videos.
\newblock In {\em Interspeech}, 2021.

\bibitem{russakovsky2015ijcv}
Olga Russakovsky, Jia Deng, Hao Su, Jonathan Krause, Sanjeev Satheesh, Sean Ma,
  Zhiheng Huang, Andrej Karpathy, Aditya Khosla, Michael Bernstein,
  Alexander~C. Berg, and Li Fei-Fei.
\newblock Image{N}et large scale visual recognition challenge.
\newblock {\em International Journal of Computer Vision}, 115(3):211--252, 12
  2015.

\bibitem{samek2021ieee}
Wojciech Samek, Gr{\'e}goire Montavon, Sebastian Lapuschkin, Christopher~J
  Anders, and Klaus-Robert M{\"u}ller.
\newblock Explaining deep neural networks and beyond: A review of methods and
  applications.
\newblock {\em Proceedings of the IEEE}, 109(3):247--278, 2021.

\bibitem{sanabria2018nips}
Ramon Sanabria, Ozan Caglayan, Shruti Palaskar, Desmond Elliott, Lo{\"\i}c
  Barrault, Lucia Specia, and Florian Metze.
\newblock How2: A large-scale dataset for multimodal language understanding.
\newblock In {\em Advances in Neural Information Processing Systems}, 2018.

\bibitem{srinivasan2019icml}
Tejas Srinivasan, Ramon Sanabria, and Florian Metze.
\newblock Analyzing utility of visual context in multimodal speech recognition
  under noisy conditions.
\newblock In {\em The How2 Challenge: New Tasks for Vision \& Language, ICML},
  2019.

\bibitem{srinivasan2020icassp}
Tejas Srinivasan, Ramon Sanabria, and Florian Metze.
\newblock Looking enhances listening: Recovering missing speech using images.
\newblock In {\em IEEE International Conference on Acoustics, Speech and Signal
  Processing}, pages 6304--6308, 2020.

\bibitem{srinivasan2020aemnlp}
Tejas Srinivasan, Ramon Sanabria, Florian Metze, and Desmond Elliott.
\newblock Fine-grained grounding for multimodal speech recognition.
\newblock In {\em Empirical Methods in Natural Language Processing}, 2020.

\bibitem{srinivasan2020bemnlp}
Tejas Srinivasan, Ramon Sanabria, Florian Metze, and Desmond Elliott.
\newblock Multimodal speech recognition with unstructured audio masking.
\newblock In {\em Workshop on Natural Language Processing Beyond Text, EMNLP},
  2020.

\bibitem{sun2016slt}
Felix Sun, David Harwath, and James Glass.
\newblock Look, listen, and decode: Multimodal speech recognition with images.
\newblock In {\em IEEE Spoken Language Technology Workshop}, pages 573--578,
  2016.

\bibitem{synnaeve2014nips}
Gabriel Synnaeve, Maarten Versteegh, and Emmanuel Dupoux.
\newblock Learning words from images and speech.
\newblock In {\em NIPS Workshop on Learning Semantics}, 2014.

\bibitem{watanabe2018interspeech}
Shinji Watanabe, Takaaki Hori, Shigeki Karita, Tomoki Hayashi, Jiro Nishitoba,
  Yuya Unno, Nelson Enrique~Yalta Soplin, Jahn Heymann, Matthew Wiesner, Nanxin
  Chen, Adithya Renduchintala, and Tsubasa Ochiai.
\newblock {ESPnet}: End-to-end speech processing toolkit.
\newblock In {\em Interspeech}, pages 2207--2211, 2018.

\bibitem{wu2021acl}
Zhiyong Wu, Lingpeng Kong, Wei Bi, Xiang Li, and Ben Kao.
\newblock Good for misconceived reasons: An empirical revisiting on the need
  for visual context in multimodal machine translation.
\newblock In {\em Association for Computational Linguistics}, 2021.

\bibitem{young2014tacl}
Peter Young, Alice Lai, Micah Hodosh, and Julia Hockenmaier.
\newblock From image descriptions to visual denotations: New similarity metrics
  for semantic inference over event descriptions.
\newblock {\em Transactions of the Association for Computational Linguistics},
  2:67--78, 2014.

\bibitem{zhou2019ijcv}
Bolei Zhou, Hang Zhao, Xavier Puig, Tete Xiao, Sanja Fidler, Adela Barriuso,
  and Antonio Torralba.
\newblock Semantic understanding of scenes through the {ADE20K} dataset.
\newblock {\em International Journal of Computer Vision}, 127(3):302--321,
  2019.

\end{thebibliography}
}

\end{document}